\documentclass[12pt, a4paper]{article}

\usepackage{amsmath}
\usepackage{amsfonts}
\usepackage{graphicx}
\usepackage{longtable}
\usepackage{color}
\usepackage{float}

\definecolor{agreen}{rgb}{0, 0.44, 0}

\newcommand{\bit}{\begin{itemize}}
\newcommand{\eit}{\end{itemize}}

\begin{document}

\title{Don’t answer the question!  How on-line Moodle-based `Data Retrieval Tests' encourage good record-keeping and a divergent experimental mindset for undergraduate physics students}

\author{P. A. Bartlett}

\date{\today}

\maketitle

\begin{abstract}
The use of Data Retrieval Tests (DRTs), as an alternative to physics laboratory notebook marking, is discussed.  The implementation of a Moodle-based, on-line DRT for 1st year physics students is described.  The advantages of using such a methodology are highlighted and student comments shown.  The paper also describes how students change their behaviour as a consequence of having an end of module DRT via `bootstrapping', both singly and in peer groups.
\end{abstract}

\section{Introduction}	

In many higher education courses, particularly Science, Technology, Engineering and Mathematics (STEM) ones, students are required to keep laboratory notebooks.  These notebooks tend to be the core of the laboratory experiment or field trip process and a data-recording methodology.  Typically, these courses are designed so that students can be trained to keep good laboratory notebooks that comply with a subject area’s ‘best practice’.  It can be quite difficult to encourage students to comply with this and so it is common for these laboratory notebooks to be assessed by academic staff, laboratory demonstrators or field-trip leaders. 

In the past, in University College London's (UCL's) physics teaching laboratories this assessment was based upon formally marking the laboratory notebook against competency rubrics. However, experience indicated that students seemed only interested in the grades associated with this process. Many students did not read the supporting feedback comments that indicated how they could improve their reporting in future.  In addition, this marking process was undertaken in such a way that every page of a laboratory notebook was formally assessed.  This was a laborious task that could take many hours for markers to complete, with no guarantee that the marker’s comments would be heeded by students.  In addition, at UCL the class numbers in the first-year teaching laboratory went from 80 in 2008 to 120 in 2013 (180 in 2018).  This imposed a significant load on this old marking system.  

\section{Data Retrieval Tests}

Due to these issues, a survey of the literature was undertaken in 2013 to identify if there were any other assessment approaches that could be implemented instead of the more traditional methods.  This resulted in the discovery of a paper \cite{DRT2010} that introduced the concept of the Data Retrieval Test.  This paper stated: ``The `Data Retrieval Test (DRT)' is an open book test in which the notebook is taken into the exam room and questions are asked on its contents. The questions are based on what the tutor knows should have been included. A survey of third year degree students showed that this type of test is seen as a fair method of assessment. It is also popular with staff as it can reduce the marking time by up to 80\%''.  

UCL students are trained on how to keep a laboratory notebook and are told that they will undertake an `open book' DRT at the end of the module.  During this test they can only use their laboratory notebook and no pens, pencils, mobile phones or internet resources can be used.  At UCL, students are given one practice experiment at the start of their first term at the university that is supported by training and a guide-document on how laboratory notebooks should be completed during an experiment.   

To ensure that students are following this guidance, each of them has a one-to-one oral assessment of their laboratory notebook twice during the term.  This gives students advice on how they are doing with their recording practice.  It also helps them to fine tune their recording techniques and gives them confidence that they will be able to undertake the DRT successfully.  It was considered that only having a DRT at the end of term would be too retrospective on its own and so these mini-orals were put in place to give students ‘in flight corrections’ to their laboratory practice. In addition, it allowed students to ask questions of experienced practitioners that they might not ask in the normal conduct of an experiment.

The DRTs initially undertaken at Edge Hill University \cite{DRT2010} and UCL were paper based.  Students would fill in the blank fields on question sheets and these could be marked manually by staff.  At UCL the DRT comprised of multiple choice, numerical and written answers that would require a complete laboratory notebook record to complete.

One thing that was immediately noticeable upon introduction of the DRT in 2013 was how it focused the minds of the students.  They knew that the DRT was coming at the end of term and because it represented a significant contribution to their course marks, they were highly motivated to make sure that they kept a good record that complied with best practice.  

It was noticed that students would come to the laboratory to work on their notebook outside allotted sessions and that they would form peer-groups to improve their practice through mutual support.  That had not been observed before the implementation of the DRTs.  This was a significant improvement as students, rather than just relying on staff to guide them, were creating communities of interest amongst their peers. The self-motivated self-improvement that the looming presence of the end of term DRT causes can be considered to be the students being made to `pull them selves up by their own bootstraps (bootstrapping). This is  done individually and in these self-help groups.

This is behaviour that UCL encourages through its ‘Connected Curriculum’ ethos where it is thought to be ‘…possible to bring university research and student education into a more connected, more symbiotic relationship…’where it is ‘…a shared ‘framework’ for thinking about how curriculum is designed, and how students can become partners in both research and educational development…’ \cite{Fung2017}.

\section{On-line Data Retrieval Tests via Moodle}

However, we at UCL have deviated away from this paper-based DRT assessment method in favour of a on-line Moodle-based approach.  Moodle is a commonly used open-source learning management system. It contains the ability to host on-line quizzes that can comprise of a range of multiple choice, ‘select the right word’, numerical and written answers.  In addition, it is possible for students to use other packages (Word or Excel, for example) to upload answers as files, into their Moodle quiz submission.  Our teaching laboratories have sufficient computers so that there is one for each student conducting the on-line DRT.  Usually these tests comprise 15 to 20 questions and the students are given 3hrs to complete it.  Examples of a multiple choice and written style questions are shown In Fig 1 and 2.

\begin{figure}[h!]
\includegraphics[width=\columnwidth]{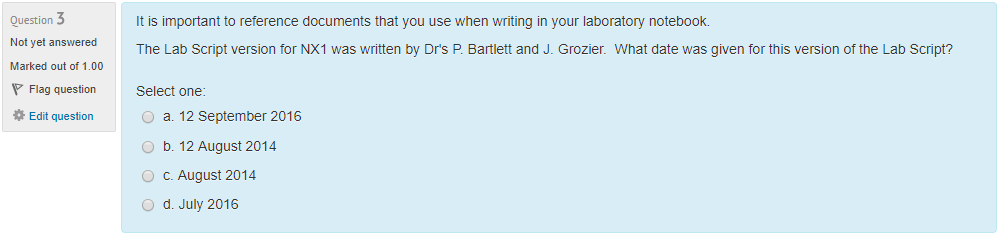}
\caption{Example of a Data Retrieval Test multiple choice question to check that students are recording important information in their laboratory notebooks}
\end{figure}

\begin{figure}[h!]
\includegraphics[width=\columnwidth]{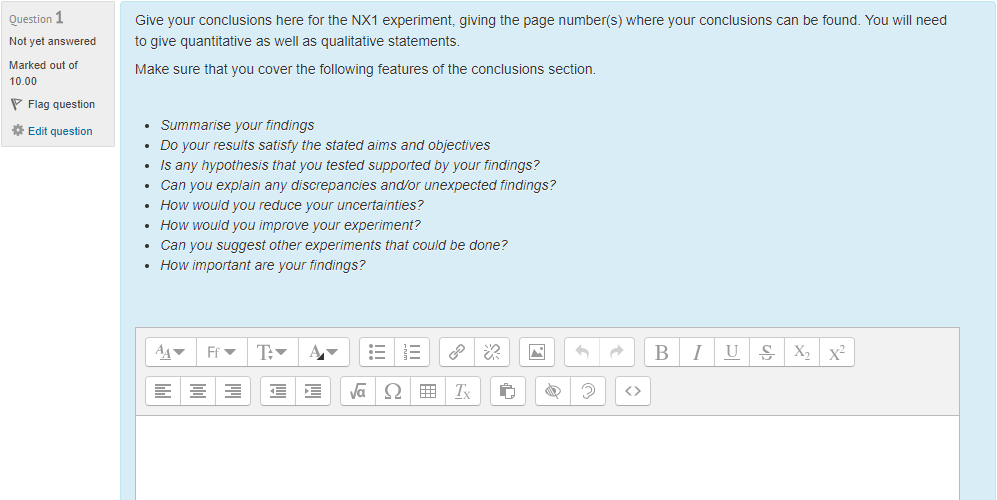}
\caption{Example of a Data Retrieval Test question that requires a written response. It tests that students have recorded information in accordance with local laboratory best practice.}
\end{figure}

For written questions, students must give the page numbers used from their notebooks, so that markers can cross reference what was written in the DRT with what is in them.  In practice, this is done by comparison via statistical sampling, but few students seem to offer written answers that do not reflect what they have in their notebooks and, occasionally, apologize for their recording deficiencies in their answer submissions (see Fig. 3).  

\begin{figure}[t!]
\includegraphics[width=\columnwidth]{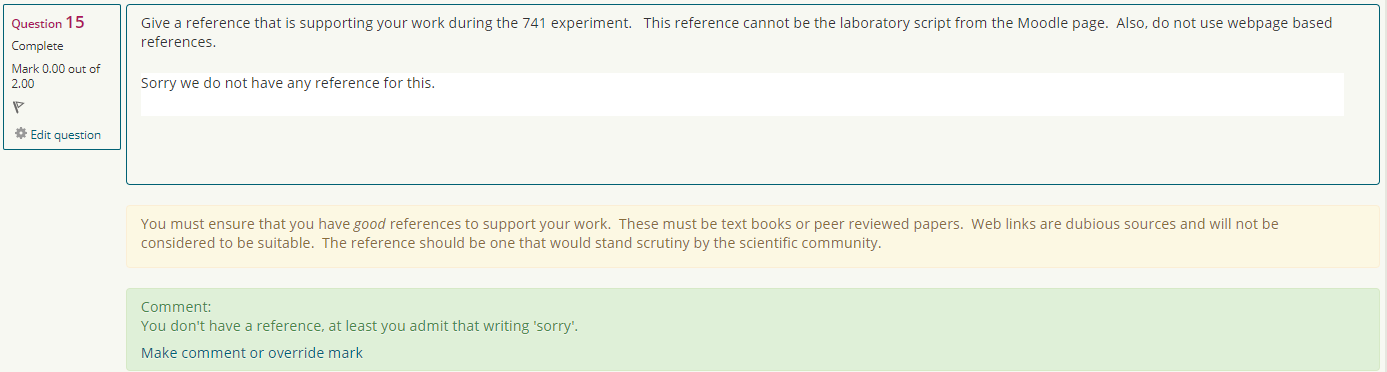}
\caption{Example of a Data Retrieval Test question that required a written answer from a student. It shows a typical response when a student does not have the required information recorded in their laboratory notebook. It also shows pre-written automated feedback (yellow field) and a comment from a marker (green field).}
\end{figure}

In the past it was noted was that the written questions posed in earlier, paper-based, iterations of the DRTs could ask students to comment on something they didn’t or couldn’t do in the laboratory.  This might be because the experiment was not a success (from their point of view) due to equipment malfunction or unexpected results or because the students deviated from the laboratory script to investigate another feature of the physics under investigation.

\section{Encouraging students to `Not answer the question'}

In the UCL physics laboratories, we encourage our students to modify what they do so they can use their creativity and enthusiasm to create their own experiments with the laboratory script just acting as a starting point.  That means that requesting that the students describe the results of an experiment that they could not do or deviated from would throw them onto turmoil as they felt they would be penalized in the DRT for not following the ‘safe’ scripted path.  To resolve this, we allow the students to NOT answer the question that is asked in the written questions.  This helps to develop the skills needed to be able to defend their work and decisions to others. In this case, those marking the DRT. The instructions the students get on the Moodle Quiz page are as follows:

``In some cases, you may find it difficult to answer a written question if you experiment did not go as planned.  In that case, explain what you did, what you found and what you think about it.  You are being assessed on what is in your laboratory notebook, so any difficulties should be recorded there''. 

Students are told that the DRT is the only time where they can answer a question that is not on in the test if they can justify this.  This seems to be a unique feature of this assessment system.  Our students have the freedom to change what they do in the laboratory and have 'permission to fail' as long as they have recorded what they did well.  They know that they can then justify NOT answering the question asked if they can justify answering one that is better suited to what they did and/or experienced.

The overarching emphasis is on using the laboratory notebook as a document of record rather than a repository of predetermined ``correct'' answers. If they can justify why they are giving a different answer that is not directly related to the question, then they will be assessed on this.  This gives considerable comfort to students that had experiments that ``went wrong'' because they can then make this the focus of their answer rather than the results that the question may have requested. 

It must be noted that the Moodle quiz system enables the automatic marking of everything but the written components of the students’ submissions. In addition, by having written questions that could be on any topic on any of the experiments, it is possible to assess that the laboratory notebook ‘best practice’ competencies are achieved without having to mark every part of the notebook itself. Both features of the Moodle-based DRT significantly reduce marking workload without reducing the testing of students’ competencies. Another advantage is that markers can do the initial marking on-line and so do not have to transport large numbers of laboratory notebooks to offices or homes to do this.

\section{Student Responses to Data Retrieval Tests}

The students were asked about what they thought of the DRT process via an end of module questionnaire.  Some representative comments are reproduced here:

``DRT itself is useful, as they help students keep a well filled in lab book, preventing them from leaving out information''.

``This is a good method which I think really works, and I appreciate the fact that it is dependent almost solely on the lab book and not revision''.

This comment is important as it also highlight that this is a test that requires no revision. 

``That (it) probably helps, (because) there are some really detailed questions like what's the brand of the apparatus or on which page the reference use in the book etc''.

``(A) good way of testing if you have sufficient detail in your lab book''.

``Data Retrieval Test is a great idea! However, the weightage is rather disproportionate''.

This reflects the high marks that are assigned to this test as part of the course grade.  It is kept high (70\%) to focus students minds regarding the importance of laboratory notebooks

``I think it is helped realise how much info I need to put in my lab book, thus making me more competent in the process''.

``I think it's quite a good idea - after all, the way the DRT is designed it's less of a test that you conduct in those three hours and more of one that you conduct across the entire term, and in that respect, I think it has achieved its objective''.

``Knowing that the Data Retrieval Test was at the end of term made me think more about writing key details down in my lab book as I know that it could be in the test'.

These comments agree with those that have been obtained through discussions with students.  They tend to think it is a good way to encourage students to keep a good laboratory notebook, but they do wish that it did not represent a significant proportion of the marks.  Which seems odd as most of their other lecture courses have an end of year exam that would represent a large proportion of the marks for that module.  

\section{Conclusions}

In conclusion, it is considered that Moodle-based DRTs are an important method for assessing student notebooks.  

Some of the benefits of using Moodle-based DRTs are as follows:

\begin{itemize}
  \item Labour intensive laboratory notebook marking can be eliminated.
  \item Non-written answer questions on Moodle can be automatically marked by Moodle itself.  This reduces the markers' marking load.
  \item Students seek to improve their own performance throughout the module via 'bootstrapping'.
  \item If students cannot answer a question (due to equipment failure or them choosing to move away from the laboratory script), they can justify this in the Moodle test and so NOT answer the question that was asked.  This applies only to written answer questions.
  \item It helps students to have the confidence to justify their work and decisions to other people (in this case, the markers).
  \item Pre-determined feedback can be given automatically by Moodle.  This is supported by comments given by markers for the written answer questions.
  \item Markers can mark on-line and so do not need to transport large numbers of notebooks to their offices or homes to mark them.
  \item It is easier for second marking to take place as the written answers for the students' DRT submissions are in one database that is accessible anywhere.
\end{itemize}

The introduction of Moodle-based Data Retrieval Tests (based on a paper-based process created by Edge Hill University in the UK) has significantly reduced the marking load of staff within UCL's 1st year physics teaching laboratories. The responsibility for keeping a good laboratory notebook has been placed squarely on the shoulders of the students. They have responded with more professional recording behaviours.  It is recommenced for other practical physics courses and for any course in a STEM practical environment.

\bibliographystyle{unsrt}
\bibliography{teachingbib,Books}

\end{document}